\documentstyle[preprint,aps,epsf]{revtex}

\newcommand{\be}{\begin{eqnarray}}
\newcommand{\ee}{\end{eqnarray}}

\begin{document}
\bibliographystyle{unsrt}
\draft
\preprint{IASSNS-HEP-98-90}

\title{High Density Quark Matter and the Renormalization Group\\
in QCD with two and three Flavors}

\author{Thomas Sch\"afer and Frank Wilczek}

\address{Institute for Advanced Study\\ 
         School of Natural Sciences\\ 
         Princeton, NJ 08540}

\maketitle

\begin{abstract}
We consider the most general four fermion operators in QCD for 
two and three massless flavors 
and study their renormalization in the vicinity 
of the Fermi surface. We show that, asymptotically, the largest 
coupling corresponds to scalar diquark condensation. Asymptotically 
the direct and iterated (molecular) instanton interactions become
equal. We provide simple arguments for the form of the operators 
that diagonalize the evolution equations. Some solutions of the flow
equations exhibit instabilities arising out of purely repulsive 
interactions.

\end{abstract}

  1. Recently there has been renewed interest in the problem 
of hadronic matter at high baryon density. It was realized early
on that at very high density, asymptotic freedom and the presence
of a Fermi surface implies that attractive interactions between
quarks cause a BCS instability, and cold quark matter is a 
color superconductor \cite{BL_84}. More recently, the problem 
was studied again, and it was emphasized that non-perturbative 
effects, instantons, could lead to gaps on the order of 100 MeV
\cite{ARW_98,RSSV_98}. A particularly interesting case is QCD
with three light flavors. In this case the preferred order parameter
involves a coupling between color and flavor degrees of freedom
\cite{ARW_98b}. The resulting coherence leads to large gaps even
for weak interactions. In addition to that, color-flavor locking
implies that chiral symmetry is broken. The usual quark condensate
is non-vanishing even in the high density phase, as is a
gauge-invariant $\Delta B = 2$ condensate, signalling true superfluidity 
\cite{ARW_98b,RSSV_98b}.

  Most of these calculations were performed in the mean field 
approximation.  (In the instanton model, a number of 
refinements have been included, see \cite{CD_98,RSSV_98b}.) 
One assumes a specific form of the interaction (instantons, one 
gluon exchange, $\ldots$), makes an {\it ansatz\/} 
for the form of the condensate, 
and solves the gap equation in some approximation. Clearly, 
it is desirable 
to analyze the structure of the theory for the 
most general form of the interaction, and to have
a guiding principle for constructing a systematic expansion 
scheme. 

 In this context, the renormalization group approach to cold
Fermi systems appears particularly promising \cite{Pol_92,Sha_95}. 
In a cold Fermi system, the only relevant interactions 
take place in vicinity of the Fermi surface. The corresponding 
excitations are quasi-particles and quasi-holes, described by an
effective action of the form
\be
 S_{eff} &=& \int dt\, d^3p\; \psi^\dagger \left( i\partial_t
 - (\epsilon(p)-\epsilon_F) \right)\psi .
\ee
We can analyze the general structure of the quasi-particle 
interactions by studying the evolution of the corresponding 
operators as we successfully integrate out modes closer and
closer to the Fermi surface. The main result of this analysis
is that, in general, four fermion, six fermion, and higher 
order interactions are suppressed as we approach the Fermi 
surface. This fixed point corresponds to Landau liquid theory 
\cite{AGD}. The only exception is the four-fermion 
operator that corresponds to two particles from opposite corners
of the Fermi surface $(p_F,-p_F)$ scattering into $(p'_F,-p'_F)$.
This kind of scattering leads to a logarithmic growth of the 
coupling constant as we approach the Fermi surface, and the 
well known BCS instability. 

  2. For QCD, this approach was first employed by Evans, Hsu, 
and Schwetz \cite{EHS_98}. These authors concentrated on the one gluon 
exchange interaction, but as we will see in a moment, their results 
are sufficient to deal with the most general quark-quark 
interaction for QCD with three (or more) flavors. We should 
emphasize here that we will restrict ourselves to massless
QCD, a spherical Fermi surface, and local operators invariant under
the appropriate chiral symmetry. 

  Following \cite{EHS_98}, we take the basic 
four fermion operators 
in $N_f=3$ QCD to be 
\be
\label{nf3_ops}
O^0_{LL} &=& (\bar\psi_L\gamma_0\psi_L)^2, \hspace{1cm}
O^0_{LR} \;=\; (\bar\psi_L\gamma_0\psi_L)(\bar\psi_R\gamma_0\psi_R) \\
O^i_{LL} &=& (\bar\psi_L\gamma_i\psi_L)^2,  \hspace{1cm}
O^i_{LR} \;=\; (\bar\psi_L\vec\gamma\psi_L)
       (\bar\psi_R\vec\gamma\psi_R) . \nonumber
\ee
Each of these operators comes in two color structures, 
for example color symmetric and color anti-symmetric 
\be
(\bar\psi^a\psi^b)(\bar\psi^c\psi^d)
  \left(\delta_{ab}\delta_{cd}\pm \delta_{ad}\delta_{bc}\right).
\ee
Nothing essentially new emerges from considering superficially 
different isospin structures, or different Dirac matrices. All 
such structures can be reduced to linear combinations of the 
basic ones (\ref{nf3_ops}), or their parity conjugates, by 
Fierz rearrangements. In total, we have to consider eight 
operators. 

  These operators are renormalized by quark-quark scattering 
in the vicinity of the Fermi surface. This means that both 
incoming and outgoing quarks have momenta $\vec p_1, \vec p_2\simeq
\pm \vec p$ and $\vec p_3, \vec p_4\simeq\pm \vec q$ with $|\vec p|, 
|\vec q| \simeq p_F$. We can take the external frequency to be zero. 
A graph with vertices $\Gamma_1$ and $\Gamma_2$ then gives \cite{EHS_98}
\be
\label{loop}
G_1 G_2 I\; (\Gamma_1)_{i'i}(\Gamma_1)_{k'k}
  \left[ -(\gamma_0)_{ij}(\gamma_0)_{kl}-\frac{1}{3}
          (\vec\gamma)_{ij}(\vec\gamma)_{kl}\right]
 (\Gamma_2)_{jj'}(\Gamma_2)_{ll'}
\ee
with $I=\frac{i}{8\pi^2}\mu^2\log(\Lambda_{IR}/\Lambda_{UV})$. 
Here $[\Lambda_{IR},\Lambda_{UV}]$ is the range of momenta that
was integrated out. We will denote the density of states on the 
Fermi surface by $N=\mu^2/(2\pi^2)$ and the logarithm of the scale
$t=\log(\Lambda_{IR}/\Lambda_{UV})$ as in \cite{EHS_98}. The 
renormalization group does not mix $LL$ and $LR$ operators, and 
it also does not mix different color structures. This means that 
the evolution equations contain at most $2\times 2$ blocks. For 
completeness, we reproduce the results of \cite{EHS_98}
\be 
\label{nf2_evol}
\frac{d(G^{LL}_0+G^{LL}_i)}{dt} &=& -\frac{N}{3}
   (G^{LL}_0+G^{LL}_i)^2 \\
\frac{d(G^{LL}_0-3G^{LL}_i)}{dt} &=& -N
   (G^{LL}_0-3G^{LL}_i)^2 \\
\frac{d(G^{LR}_0+3G^{LR}_i)}{dt} &=& 0\\
\frac{d(G^{LR}_0-G^{LR}_i)}{dt} &=& -\frac{2N}{3}
   (G^{LR}_0-G^{LR}_i)^2
\ee
In this basis the evolution equations are already diagonal. The coupling
$G_1=G^{LL}_0+G^{LL}_i$ evolves as 
\be
 G_1(t) &=& \frac{1}{1+(N/3)G_1(0)t}
\ee
with analogous results for the other operators. Note that the evolution
starts at $t=0$ and moves towards the Fermi surface $t\to-\infty$.
If the coupling is attractive at the matching scale, $G_1(0)>0$, it 
will grow during the evolution, and reach a Landau pole at $t_c=3/(N
G_1(0))$. The location of the pole is controlled by the initial value 
of the coupling and the coefficient in the evolution equation. If the 
initial coupling is negative, the coupling decreases during the 
evolution. The second operator in (\ref{nf2_evol}) has the largest 
coefficient and will reach the Landau pole first, unless the initial 
value is very small or negative. In this case, subdominant operators 
may dominate the pairing.  

  The particular form of the operators that diagonalize the 
evolution equations is easily understood. Let us first order 
the operators according to the size of the coefficient in the 
evolution equations
\be
\label{O_dom}
O_{dom} &=& (\bar\psi_L\gamma_0\psi_L)^2 - 
               (\bar\psi_L\vec\gamma\psi_L)^2,\\
O_{sub,1} &=& (\bar\psi_L\gamma_0\psi_L)(\bar\psi_R\gamma_0\psi_R)  
  -\frac{1}{3} (\bar\psi_L\vec\gamma\psi_L)(\bar\psi_R\vec\gamma\psi_R),\\
O_{sub,2} &=& (\bar\psi_L\gamma_0\psi_L)^2 
  + \frac{1}{3} (\bar\psi_L\vec\gamma\psi_L)^2,\\
\label{O_mar}
O_{mar}   &=& (\bar\psi_L\gamma_0\psi_L)(\bar\psi_R\gamma_0\psi_R)  
  + (\bar\psi_L\vec\gamma\psi_L) (\bar\psi_R\vec\gamma\psi_R).
\ee
This result can be made more transparent by Fierz rearranging the 
operators. We find
\be
O_{dom} &=&   2(\psi_LC \psi_L)(\bar\psi_L C\bar\psi_L),\\
O_{sub,1} &=& \frac{1}{3}(\psi_L C\vec\gamma\psi_R)
 (\bar\psi_RC\vec\gamma\psi_L) + \ldots ,\\
O_{sub,2} &=& \frac{4}{3}(\psi_L C\vec\Sigma\psi_L)
 (\bar\psi_LC\vec\Sigma \bar\psi_L) ,\\
O_{mar}   &=&  \frac{1}{2}(\psi_LC\gamma_0\psi_R)
 (\bar\psi_RC\gamma_0\psi_L) + \ldots . 
\ee
This demonstrates that the linear combinations in 
(\ref{O_dom}-\ref{O_mar}) correspond to simple structures in the 
quark-quark channel. It also means that it might have been more 
natural to perform the whole calculation directly in a basis of diquark 
operators.  

 The full structure of the $(LR)$ operators is $O_{sub,1},O_{mar}=
(\psi C\gamma\tau_{S,A}\psi)(\bar\psi C\gamma\tau_{S,A}\bar\psi)+
(\psi C\gamma\gamma_5\tau_{A,S}\psi)(\bar\psi C\gamma\gamma_5
\tau_{A,S}\bar\psi)$ where $\tau_{S,A}$ are symmetric/anti-symmetric
isospin generators. Note that because the two structures have different
flavor symmetry, the flavor structure cannot be factored 
out.

 The dominant operator corresponds to pairing in the scalar diquark 
channel, while the subdominant operators contain vector diquarks. 
Note that from the evolution equation
alone we cannot decide what the preferred color channel is. 
To decide this question, we must invoke 
the fact that ``reasonable'' interactions,
like one gluon exchange, are attractive in the color anti-symmetric
repulsive in the color symmetric channel. Indeed, it is the color
anti-symmetric configuration that minimizes the total color flux
emanating from the quark pair.  If the color wave
function is anti-symmetric, the dominant operator fixes the 
isospin wave function to 
be anti-symmetric as well. 

The dominant operator does not distinguish 
between scalar and pseudoscalar diquarks.  Indeed, 
for $N_f \geq 3$ the basic 4-quark operators  
consistent with chiral symmetry also exhibit an accidental
axial baryon symmetry, under which scalar and pseduscalar diquarks are
equivalent. For $N_f=3$ this 
degeneracy is lifted by (formally irrelevant) six-fermion operators 
\cite{RSSV_98b}.  The form of the dominant operator indicates the 
existence of potential instabilities, but does not itself indicate 
how they are resolved.  For example, to see whether color-flavor 
locking \cite{ARW_98b} is important we should use the operator as 
input to a variational calculation. 

  3. For two flavors, we have to take into account additional 
operators. At first hearing it might seem odd that with fewer basic 
entities we encounter more basic operators. It occurs because for 
$N_f =2$, but not for larger values, two quarks of the same chirality 
can form a chiral $SU(2) \times SU(2)$ singlet. Related to this, for 
$N_f=2$ we have additional $U(1)_A$ violating four fermion operators.  
These operators are induced by instantons. 

For $N_f=3$,
instantons are six fermion operators, and they are 
irrelevant in the technical sense.  This does not mean they are
physically irrelevant, particularly since they break a residual symmetry.
The formation of the gap will cause the evolution of the couplings
to stop, and the instanton coupling remains at a finite value. 
Instantons have important physical effects, even for $N_f=3$ 
\cite{RSSV_98b}. Most notably, instantons cause quark-antiquark
pairs to condense (even in the high density phase), and lift the 
degeneracy between the scalar and pseudoscalar diquark condensates. 

The new operators are 
\be
\label{inst_ops}
 O_S &=& \det_f (\bar\psi_R\psi_L), \hspace{1cm}
 O_T \;=\; \det_f (\bar\psi_R\vec\Sigma\psi_L)
\ee
Both operators are determinants in flavor space. For quark-quark
scattering, this implies that the two quarks have to have different
flavors. The fact that the flavor structure is fixed implies that 
the color structure is fixed, too. For a given $(qq)$ spin, only
one of the two color structures contributes. Finally, both quarks
have to have the same chirality, and the chirality is flipped 
by the interaction. 

 These considerations determine 
the structure of the evolution equations (see Fig. 1). 
Two left handed quarks can interact via one of the instanton operators,
become right handed, and then rescatter through an anti-instanton, 
or through one of the $U(1)_A$ symmetric $RR$ operators. The result 
will be a renormalization of the $LL$ vertex in the first case, and 
a renormalization of the instanton in the second. The iterated 
instanton-anti-instanton interaction was discussed in great detail 
in \cite{SSV_95}, and was argued to play an important role in the 
high temperature \cite{SSV_95} and high density phase of QCD 
\cite{RSSV_98b}. We should note that the flavor structure will 
always remain a determinant. Even though instantons generate all 
the Dirac structures in (\ref{nf3_ops}), the color-flavor structure 
is more restricted.

Evidently, 
instantons do not affect the evolution of the $LR$ couplings at all. 
The evolution equations of the $LL$ couplings are modified to become 
(henceforth we drop the  
subscript $LL$ from): 
\be
\label{nf3_evol}
\frac{dG_0}{dt} &=& \frac{N}{2} \Bigg\{
  -G_0^2 + 2G_0G_i - 5G_i^2 - K_S^2 + 2K_S K_T - 5 K_T^2 \Bigg\} \\
\frac{dG_i}{dt} &=& \frac{N}{2} \Bigg\{
 \frac{1}{3}G_0^2 - \frac{10}{3}G_0G_i + \frac{13}{3}G_i^2 
 +\frac{1}{3}K_S^2 - \frac{10}{3} K_S K_T + \frac{13}{3}K_T^2 \Bigg\} \\
\frac{dK_S}{dt} &=& \frac{N}{2} \Bigg\{
  2\left(-G_0 + G_i \right)K_S 
 +2\left( G_0 - 5 G_i \right)K_T \Bigg\} \\
\frac{dK_T}{dt} &=& \frac{N}{2} \Bigg\{
  \frac{2}{3}\left( G_0 - 5 G_i \right)K_S 
 +\frac{2}{3}\left( -5G_0 +13 G_i \right)K_T \Bigg\}
\ee
These equations can be decoupled as
\be
\label{nf3_evol_2}
\frac{dG_1}{dt} &=& - \frac{N}{3} \left( G_1^2+K_1^2 \right), \\
\frac{dK_1}{dt} &=& - \frac{2N}{3}\; G_1 K_1, \\
\frac{dG_2}{dt} &=& -  N \left( G_2^2+K_2^2 \right), \\
\frac{dK_2}{dt} &=& - 2N \; G_2 K_2 , 
\ee
where $G_1=G_0+G_i, K_1=K_S+K_T$ and $G_2=G_0-3G_i, K_2=K_S-3K_T$.
The equations for $G,K$ decouple even further. We have
\be
\label{nf3_evol_3}
\frac{d(G_2+K_2)}{dt} &=& -N \left( G_2 +K_2 \right)^2 \\
\frac{d(G_2-K_2)}{dt} &=& -N \left( G_2 -K_2 \right)^2 ,
\ee
as well as the analogous equation for $G_1,K_1$. These differential
equations are now trivial to solve, leading to
\be
\label{nf3_flow}
 G_2(t) &=& \frac{1}{2}\left\{ \frac{1}{a+Nt} + \frac{1}{b+Nt}\right\}, \\
 K_2(t) &=& \frac{1}{2}\left\{ \frac{1}{a+Nt} - \frac{1}{b+Nt}\right\},
\ee
again with the analogous result holding for $G_1,K_1$. Here, 
$a,b=(G_2(0)\pm K_2(0))^{-1}$. The result implies that $G_2$ 
and $K_2$ will grow and eventually reach a Landau pole if either 
$a$ or $b$ is positive. The location of the pole is determined
by the smaller of the values, $t_c=-a/N$ or $t_c=-b/N$. The 
same is true for $G_1$ and $K_1$, but the couplings evolve 
more slowly, and the Landau pole is reached later. 

At this level a number of qualitatively 
different scenarios are possible, depending on the 
sign and relative magnitude of $G(0)$ and $K(0)$, see Fig. 2 (henceforth 
we drop all subscripts). If $G(0)$ and $K(0)$ 
are both positive then they will both grow, and the location of the 
nearest
Landau pole is determined by $G(0)+K(0)$.  The asymptotic ratio of 
the two couplings is 1. 
If $G(0)$ and $K(0)$ are both negative, and the 
magnitude of $G(0)$ is bigger than the magnitude of $K(0)$, then 
the evolution drives both couplings to zero. These are the 
standard cases.  Attraction leads to an instability, and repulsive
forces are suppressed. 

More interesting cases arise when the sign of the two couplings is
different.  The case $G(0),K(0)<0$ and 
$|K(0)|>|G(0)|$ is especially weird. 
Both $G(0),K(0)$ are repulsive, but the evolution 
drives $G(0)$ to positive
values.  Both couplings reach a Landau pole, and near the pole 
their asymptotic 
ratio approaches minus one. Similarly, we can have a negative $G(0)$ and 
positive $K(0)$ with $K(0)>|G(0)|$. Again, the evolution will drive 
$G(0)$ to positive values.

 4.  The dominant 
and sub-dominant instanton operators are 
\be
O_{dom} &=&\det_f\left[ (\bar\psi_R\psi_L)^2 - 
  (\bar\psi_R\vec\Sigma\psi_L)^2 \right],\\
O_{sub} &=& \det_f\left[(\bar\psi_R\psi_L)^2 +\frac{1}{3} 
  (\bar\psi_R\vec\Sigma\psi_L)^2\right].
\ee
Upon Fierz rearrangement, we find
\be
O_{dom} &=& 2(\psi_LC\tau_2\psi_L)(\bar\psi_R C\tau_2\bar\psi_R), \\
O_{sub} &=& \frac{2}{3}(\psi_LC\tau_2\vec\Sigma\psi_L)
 (\bar\psi_R C\tau_2\vec\Sigma\bar\psi_R),
\ee
corresponding to scalar and tensor diquarks. Both operators are
flavor singlet. Overall symmetry then fixes the color wave functions,
anti-symmetric $\bar 3$ for the scalar, and symmetric 6 for the tensor.
The dominant pairing induced by instantons is in the scalar diquark
channel, the only other attractive channel is the tensor. 
All this neatly confirms the scenario discussed in \cite{ARW_98,RSSV_98}.
Note that 
condensation in the tensor channel violates rotational symmetry. As
a result, the gap equation has additional suppression factors and
the gap is very small \cite{ARW_98}. 

Just as we found for $N_f \geq 3$, there is an appealing heuristic
understanding for the amazingly simple behavior of the evolution
equations, obtained by focussing on the diquark channels.  
Instantons distinguish between scalar diquarks with positive and 
negative parity. $G+K$ corresponds to the positive parity operator 
$(\psi C\gamma_5\psi)$ and $G-K$ to the negative parity 
$(\psi C \psi)$.  The asymptotic approach of $G/K\to 1$, then 
corresponds to the fact that scalar  
diquark condensation is favored over pseudoscalar diquark 
condensation. This is always the case if $K(0)>0$. We also 
understand the strange case $G(0),K(0)<0$ and $|K(0)|>|G(0)|$.
In this case the interaction for scalar diquarks is repulsive,
but the interaction in the pseudoscalar channel is attractive
and leads to an instability. Note that this can only happen
if we have the ``wrong'' sign of the instanton interaction, {\it
i.e}. for $\theta = \pi$. 
Similarly, we can understand why the asymptotic ratio of the molecular
(instanton-anti-instanton) and direct instanton couplings approaches
$G/K=\pm 1$. Instantons induce a repulsive interaction
for pseudoscalar diquarks. During the evolution, this coupling
will be suppressed, whereas the attractive scalar interaction 
grows. But this means that in the pseudoscalar channel, the repulsive 
(instanton) and attractive (molecular) forces have to cancel in the 
asymptotic limit, so the effective couplings become equal. 

 We have not made an attempt to match all the coupling constants
to a realistic model at the UV scale. To do so would require an 
understanding of the instanton density, the relevant value 
of $\alpha_s$, the screening mechanism, and many other things. 
{}From the form of the instanton vertex we can fix the ratio 
of the two instanton-like couplings, $K_T/K_S=1/(2N_c-1)$
\cite{SVZ_80b,SS_98}. This again shows that the tensor channel
is not expected to be important. Also, both one gluon exchange
and higher order instanton effects give $G^{LL}_0>0,-G^{LL}_i>0$.
Thus the favored 
scenario is that both instanton and $U(1)_A$ symmetric 
couplings flow at the same rate, and pairing is dominated by scalar 
diquarks. 

 5. In summary, 
we find that the renormalization group analysis broadly supports the
findings of \cite{ARW_98,RSSV_98}.  The dominant coupling corresponds
to scalar diquark condensation, the sub-dominant coupling to tensor 
diquarks. Asymptotically, $U(1)_A$ breaking and $U(1)_A$ symmetric
couplings flow at the same rate. Since instantons have a definite
isospin structure, there is one flavor symmetric coupling that 
evolves independently of instantons. Asymptotically, this 
coupling also flows at the same rate. In principle there is  
the possibility of flavor and color symmetric diquark 
condensates, but none of the model interactions so far proposed 
is attractive in that channel. 

The analysis presented here is incomplete in several ways. 
Realistic interactions
(one-gluon exchange, instantons, $\ldots$) are momentum dependent, and 
that should be included in the evolution. This complication 
is particularly
important for one-gluon exchange, because in perturbation theory 
the diagram is not completely screened, and has a divergence for
small momentum transfers. For $N_f=3$ a
self-consistent calculation ought to be possible, 
because color-flavor locking completely screens the interaction. 

In any case, the renormalization group only determines the running
of the couplings from the matching point down to some infrared
scale. If a subdominant coupling is unusually large at the 
matching scale, it might still dominate the pairing. 
To decide the form of the pairing and determine the gap, 
given the couplings, still requires a 
variational calculation along the lines of \cite{ARW_98b,RSSV_98b}.

Acknowledgments: After this work was begun, we learned that 
the authors of \cite{EHS_98} have also extended their study 
to include instanton operators.
This work was supported in part by NSF-PHY-9513835.

\newpage
\bibliography{rev}


\newpage\noindent
\begin{figure}
\caption{\label{fig_rge}
Chiral structure of the evolution equations. }
\end{figure}

\begin{figure}
\caption{\label{fig_evolve}
Solutions to the evolution equations for cases with
different relative strength of the $U(1)_A$ violating
interaction $K$ (dashed line) and $U(1)_A$ conserving 
interaction $G$ (solid line). }
\end{figure} 

\setcounter{figure}{0}

\begin{figure}
\begin{center}
\vspace{4cm}
\epsfxsize=16cm
\epsffile{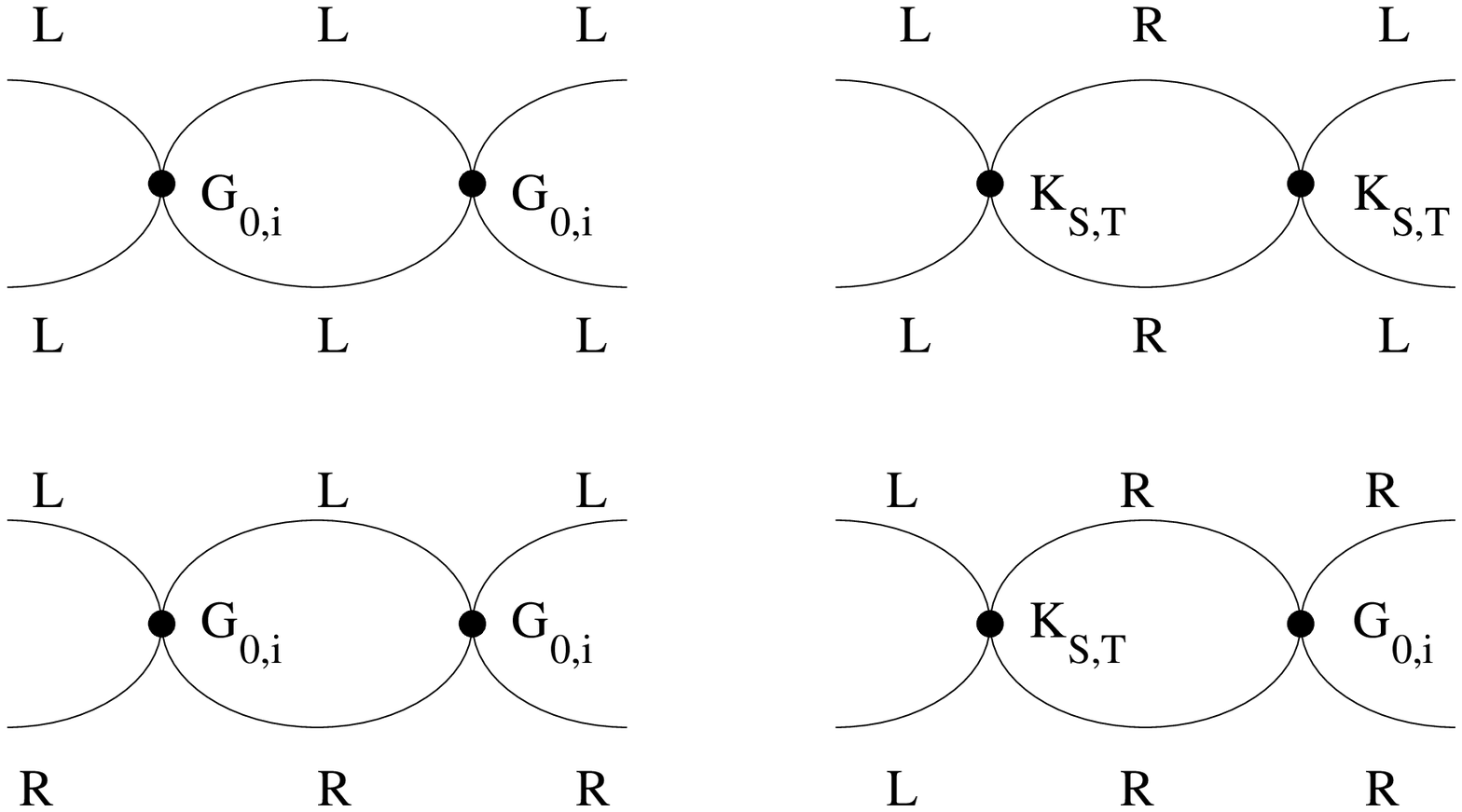}
\end{center}
\caption{}
\end{figure}

\begin{figure}
\begin{center}
\vspace*{-4cm}
\epsfxsize=16cm
\epsffile{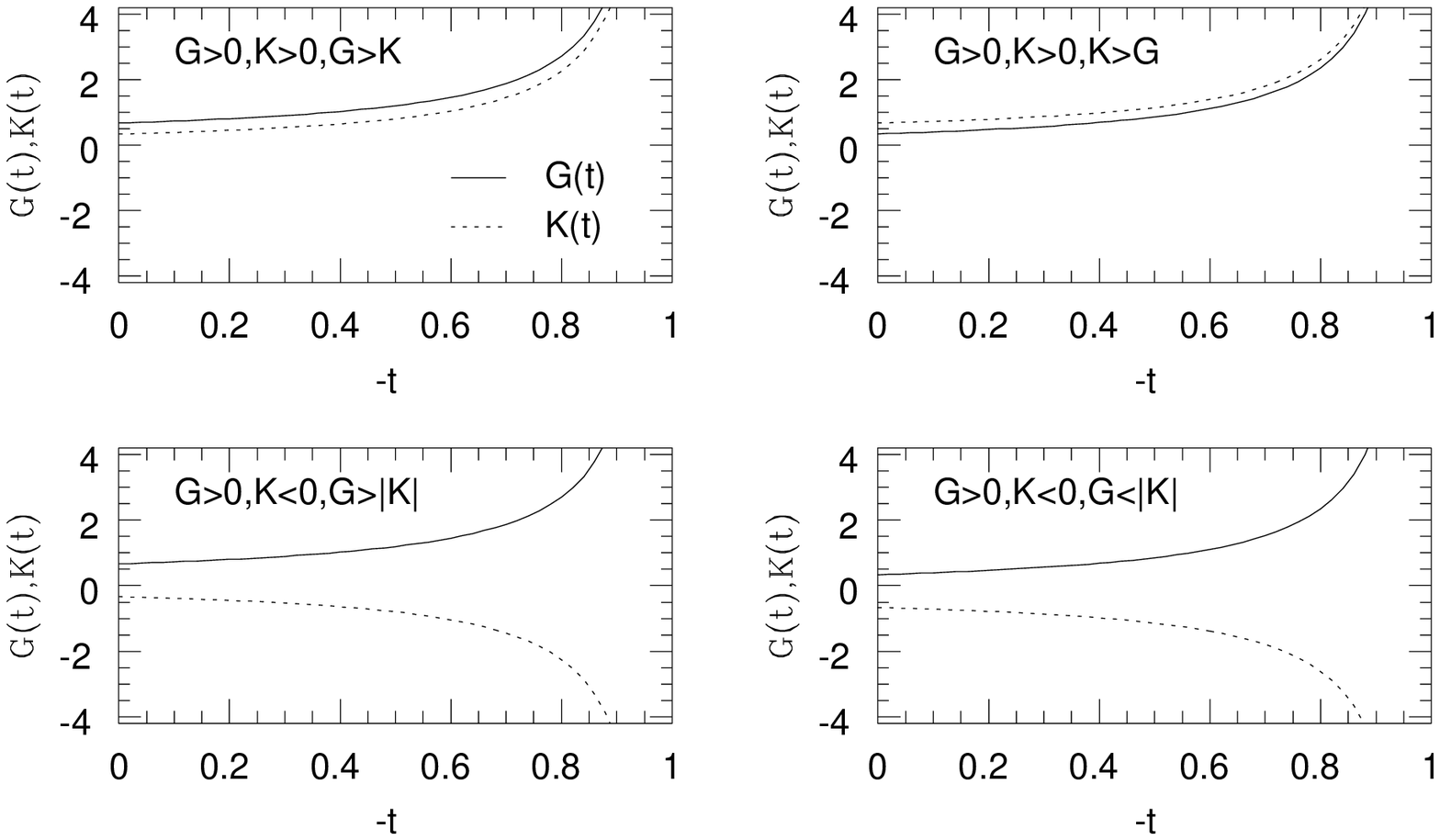}
\vspace*{-6cm}
\epsfxsize=16cm
\epsffile{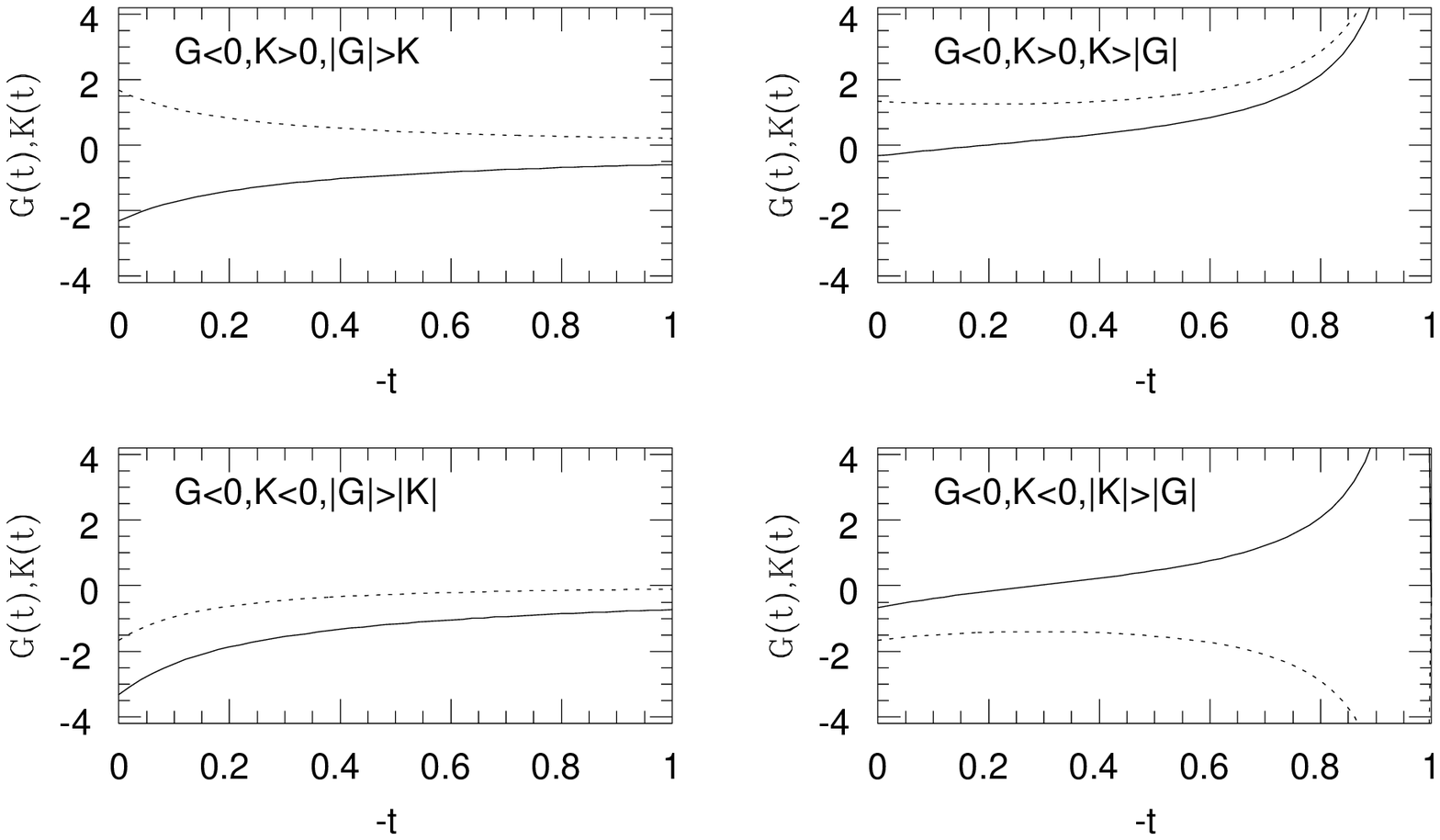}
\vspace*{-4cm}
\end{center}
\caption{}
\end{figure}

\end{document}